\documentstyle[12pt]{article}

\newcommand{\be}{\begin{equation}}
\newcommand{\ee}{\end{equation}}
\newcommand{\bea}{\begin{eqnarray}}
\newcommand{\eea}{\end{eqnarray}}

\newcommand{\de}{\partial}
\def\esp#1{e^{#1}}
\newcommand{\AAA}{{{\cal A}(G)}}
\def\Zint{{Z \kern -.45 em Z}}
\def\complex{{\kern .1em {\raise .47ex \hbox
{$\scriptscriptstyle |$}}
\kern -.4em {\rm C}}}
\def\real{{\vrule height 1.6ex width 0.05em depth 0ex
\kern -0.06em {\rm R}}}
\def\bra#1{\langle #1 |}
\def\ket#1{| #1\rangle }
\def\MPL #1 #2 #3 {{\sl Mod.~Phys.~Lett.}~{\bf#1} (#3) #2}
\def\NPB #1 #2 #3 {{\sl Nucl.~Phys.}~{\bf B#1} (#3) #2}
\def\PLB #1 #2 #3 {{\sl Phys.~Lett.}~{\bf B#1} (#3) #2}
\def\PR #1 #2 #3 {{\sl Phys.~Rep.}~{\bf#1} (#3) #2}
\def\PRD #1 #2 #3 {{\sl Phys.~Rev.}~{\bf D#1} (#3) #2}
\def\PRL #1 #2 #3 {{\sl Phys.~Rev.~Lett.}~{\bf#1} (#3) #2}
\def\RMP #1 #2 #3 {{\sl Rev.~Mod.~Phys.}~{\bf#1} (#3) #2}
\def\ZPC #1 #2 #3 {{\sl Z.~Phys.}~{\bf C#1} (#3) #2}
\def\IJMP #1 #2 #3 {{\sl Int.~J.~Mod.~Phys.}~{\bf#1} (#3) #2}

\begin{document}

\vspace*{1cm}
\begin{center}
  \begin{Large}
  \begin{bf}
Algebraic Treatment of Compactification  on Noncommutative Tori\\
  \end{bf}
  \end{Large}
\end{center}
  \vspace{5mm}
\begin{center}
  \begin{large}
R. Casalbuoni\footnote{On leave from
Dipartimento di Fisica Universit\`a di Firenze,
I-50125 Firenze, Italia}\\
  \end{large}
D\'epartement de
Physique Th\'eorique, Universit\'e de Gen\`eve\\ CH-1211 Gen\`eve
4, Suisse \\{\tt{e-mail: CASALBUONI@FI.INFN.IT}}\\
\end{center}
\vspace{0.5cm}
\begin{quotation}
\begin{center}
  \begin{bf}
  ABSTRACT
  \end{bf}
\end{center}
  \vspace{0.5cm}

\noindent
In this paper we study the compactification conditions of
the M theory on $D$-dimensional noncommutative tori. The
main tool used for this analysis is the algebra ${\cal
A}(Z^D)$ of the projective representations of the
abelian group $Z^D$. We exhibit the explicit solutions
in the space of the multiplication algebra of ${\cal
A}(Z^D)$, that is the algebra generated by right and left
multiplications.

\end{quotation}
\vspace{2.5 cm}
\begin{center}
UGVA-DPT 1998/01-996
\end{center}
\vspace{1cm}
\newpage

\section{Introduction}

In the last few years there has been a renewed interest in
string theories mostly motivated by the discovery of string
dualities \cite{duality}. This fact has induced to conjecture about
the existence of a still unknown M theory, which is supposed to
underly the known superstring models. A candidate theory has
been proposed in \cite{BFSS}. This amounts to a description,
for $N\to\infty$, of $N$ interacting $D0$ branes
\cite{Polchinski}, that is branes on which strings can end.
A completely new feature is that the $D0$ branes are
described by coordinates which are valued in the space of
the $N\times N$ hermitian matrices \cite{witten}. Since M
theory is supposed to describe gravity, one should be able
to derive from it the conventional space-time structure.
Therefore, given the fact that the $D0$ branes live in  10
dimensions, the usual space-time description
will arise only after compactification along  various space
directions. Among different possibilities, a certain
attention has been given to compactifications on circles and
on tori, because these give rise in a natural way to strings
and membranes. However, the intrinsic non commuting nature
of the coordinates of the $D0$ branes leads to
compactification on noncommutative geometries, such as
noncommuting tori \cite{BFSS,taylor,schwarz,cinesi}.

In this paper we will present an algebraic study of the non
commutative torus based on the use of projective
representations of the abelian group $Z^D$. These
representations form a noncommutative algebra ${\cal
A}(Z^D)$  which  can be represented
on the space of  the related multiplication algebra
generated by right and left multiplications of ${\cal
A}(Z^D)$.
Using this fact we will be able
to get explicit expressions for the compactified coordinates
in terms of a particular set of derivations on the algebra
and identifying the gauge field part of the coordinates with
the left multiplications. This follows from the simple
observation that the associativity  requires
that left and right
multiplications commute.

A solution to the problem presented here has been given in
ref. \cite{schwarz} in the framework of the Connes
formulation of noncommutative geometry \cite{connes}. But
our  treatment  determines in a unique way the
derivation part of the compactified coordinates, and allows
us to give in explicit terms the more general realization of
the gauge part using relatively simple algebraic techniques.
In particular we emphasize the relevance of the projective
representations  for noncommuting compactifications in the
framework of the M theory. In fact these techniques can be
easily extended to noncommuting geometries more involved
than the one of the noncommuting torus.

\section{Compactification on noncommuting tori}

$D0$ branes are point-like objects described by $N\times N$
hermitian matrices, $X^\mu_{i_1,i_2}$, $\mu= 1,\cdots, 9$,
$i_1,i_2=1,\cdots N$, moving in a space-time $\real^9\times
\real$ \cite{BFSS}. Since the usual space-time description
is supposed to arise from this theory, several $D0$ brane
coordinates nee to be compactified. We will study here the
compactification on a noncommuting $D$-dimensional  torus,
$T^D$, ($D<9$). Since $X^\mu$ are dynamical variables one
cannot require  directly that they describe a torus
geometry. The problem can be solved along the lines outlined
in ref. \cite{compact}. That is, by observing that $T^D$ is
given by the quotient $R^D/Z^D$, where $Z$ is the group of
the integers, one can describe the motion of the $D0$ branes
on $R^D$, and then take the quotient with respect to the
group $Z^D$. Technically this is done by taking infinite
copies of the $D0$ branes through the  extension of the
matrices $X^\mu_{i_1,i_2}$ to $X^\mu_{(i_1,a_1)(i_2,a_2)}$,
where $a_1$ and $a_2$ are elements of $Z^D$, that is of the
form
\be
\sum_{\mu=1}^D m^\mu e_{(\mu)}
\label{vettori}
\ee
where $m^\mu\in Z$, and $e_{(\mu)}$ are  a set of linearly
independent vectors defining the space lattice. Then we go
to the quotient by requiring the theory to be invariant with
respect to the compactification condition
\be
U(a)^{-1} X^\mu U(a)=X^\mu+a^\mu
\label{compactification}
\ee
where $a^\mu$ are the components of the vector $a$  (see eq.
(\ref{vettori})), and  $U(a)$ are unitary operators. The
operators $U(a)$ act on the group indices of $X^\mu$, and by
consistency they must belong to a projective representation
of $Z^D$
\be
U(a)U(b)=\esp{i\alpha(a,b)} U(a+b)
\ee
If the co-cycle $\alpha(a,b)$ is trivial, then  we have
vector representations and we speak of a commutative torus,
otherwise we are in the noncommutative case. The
compactification condition (\ref{compactification}) tells us
that the operators $U$, acting on group indices, belong to
the regular projective representation of the group $Z^D$.
Therefore, the mathematical problem of compactification on a
noncommutative torus $T^D$ is now completely defined, and it
consists in describing the projective regular representation
of $Z^D$, and in finding operators $X^\mu$ on this space,
such as to satisfy eq. (\ref{compactification}).

To realize this program, we start by considering  projective
representations of an abelian group $G$, which will be
specialized, later on, to $Z^D$. Let us take an arbitrary
projective representation of the group $G$. This defines an
associative noncommutative algebra $\AAA$
\be
x(a)x(b)=\esp{i\alpha(a,b)}x(a+b)=\sum_{c\in
G}f_{abc}x(c),~~~~~a,b\in G,~~~x(a),x(b)\in \AAA
\ee
where  $f_{abc}=\delta_{a+b,c}\esp{i\alpha(a,b)}$ are the
structure constants of the algebra. The associativity
requires the phase $\alpha(a,b)$ to satisfy the co-cycle
condition
\be
\alpha(a,b)+\alpha(a+b,c)=\alpha(b,c)+\alpha(a,b+c)
\ee
It is not difficult to show that the co-cycle is   an
antisymmetric bilinear mapping $G\times G\to \real$. The
regular representation can be evaluated in terms of the
right and left multiplications on  $\AAA$. To this end it is
convenient to introduce vectors $\bra x$ with components
$x(a)$ ($\bra x_a=x(a)$), $a\in G$, and the corresponding kets $\ket x$
\cite{casalbuoni}. We define
\be
R_a |x\rangle=|x\rangle x(a),~~~~\bra
{x}L_a=x(a)\bra{x}~~~x(a)\in{\AAA}
\label{eigenequation}
\ee
In the following we will use also $L_a^T\ket x=x(a)\ket x$.
The matrices of the right and left multiplications can be
expressed in terms of the structure constants of the
algebra, and one can easily show that, due to the
associativity, the left and right multiplications give a
representation of the algebra itself
\be
R_aR_b=\sum_{c\in G}f_{abc}R_c,~~~~L_aL_b=\sum_{c\in G}f_{abc}L_c
\ee
and that
\be
[R_a,L_b^T]=0
\label{commutativity}
\ee
These matrices can be expressed in terms of the structure
constants, obtaining
\be
(R_a)_{bc}=f_{bac}=\delta_{a+b,c}\esp{i\alpha(b,a)}, ~~~~
(L_a)_{bc}=f_{acb}=\delta_{a+c,b}\esp{i\alpha(a,c)}
\ee
It follows that we can identify the operators $U(a)$ with,
for instance, the matrices $R_a$, for $G=Z^D$. The second
step of our problem is to find out the operators $X^\mu$.
This is easily solved by introducing a derivation $D$ on the
algebra, that is a linear mapping satisfying the Leibnitz
rule. We will call $d$, the matrix of this application, that
is $Dx(a)=\sum_{b\in G}d_{ab}x(b)$. Acting with a derivation
upon the first of equations (\ref{eigenequation}), one can
prove the following identity
\be
R(x(a))^{-1}dR(x(a))=d-R(x(a))^{-1}R(Dx(a))]
\ee
where we have used the more explicit notation $R(x(a))\equiv
R_a$.  Then, a particular solution to the compactification
condition is given by operators $D^\mu$ such that
\be
D^\mu x(a)=-a^\mu x(a),~~~~d^\mu_{ab}=-a^\mu\delta_{a,b}
\ee
It can be checked immediately that these operators are
indeed derivations (that is they satisfy the Leibnitz rule).
This is not the most general solution, since we can always
add to $d^\mu$ any operator commuting with $U(a)$. Equation
(\ref{commutativity}) gives us such a set of operators.
Then, the general solution to eq. (\ref{compactification})
is given by
\be
X^\mu=d^\mu +{A}^\mu
\ee
where ${A}^\mu$ is an arbitrary linear combinations of $L_a^T$
\be
{A}^\mu=\sum_{a\in Z^D}f_a^\mu L_a^T
\ee
Notice that the operators $L_a^T$ define the so called
opposite algebra
\be
L_a^TL_b^T=\esp{-i\alpha(a,b)}L_{a+b}^T
\ee
By introducing the algebra valued quantities
\be
f^\mu=\sum_{a\in Z^D}f_a^\mu x(a), ~~~~f^\mu\in{\cal
A}(Z^D)
\ee
we can write
\be
X^\mu=d^\mu+L_{f^\mu}^T
\ee
These operators can be considered as connections with a
curvature given by
\be
[X^\mu,X^\nu]=-L_{F^{\mu\nu}}^T
\ee
where
\be
F^{\mu\nu}=D^\mu f^\nu-D^\nu f^\mu-[f^\mu,f^\nu]\in {\cal
A}(Z^D)
\label{curvature}
\ee
and we have used
\be
[L(x(a))^T,d]=L(Dx(a))^T
\ee
In the case
of $D=1$ and of the commuting $D$-torus, the representations of 
the algebra
are one-dimensional and they are given by the characters
\be
x(m^\mu e_{(\mu)})=\chi_{\vec q}(\vec m)=\esp{i\vec m\cdot \vec q}
\label{esponenziale}
\ee
where $0\le q_\mu\le 2\pi$ are the coordinates on the torus $T^D$.
In this case, the derivation introduced before is essentially $\de/\de
q_\mu$. From this point of view it is interesting to notice
that one could retain the parameterization of eq.
(\ref{esponenziale}) also in the case of a non commutative
$D$-torus, by requiring
\be
[q_\mu,q_\nu]=i\epsilon_{\mu\nu}
\ee
with $\epsilon_{\mu\nu}$ commuting with $q_\mu$.
In fact from
\be
\esp{im^\mu q_\mu}q_\nu\esp{-im^\mu q_\mu}=q_\nu-m^\mu
\epsilon_{\mu\nu}
\ee
we get
\be
\esp{im^\mu q_\nu}\esp{in^\nu q_\nu}\esp{-im^\mu q_\mu}=
\esp{-i m^\mu n^\nu\epsilon_{\mu\nu}}\esp{in^\nu q_\nu}
\ee
allowing us to make the following identifications
\be
x(a)=\esp{im^\mu q_\mu},~~~~~
\alpha(a,b)=-m^\mu n^\nu\epsilon_{\mu\nu}
\ee
where $m^\mu$ and $n^\mu$ are the components of $a$ and $b$.
Therefore, one could think to characterize the
representations of the projective algebra by the
noncommuting operators $\vec q$, writing
\be
x(a)\equiv x_{\vec q}(\vec a)
\ee
Then, the generic element of the algebra $\AAA$
\be
\tilde f(\vec q)=\sum_{a\in Z^D}f(\vec a) x_{\vec q}(\vec a)
\label{NC-Fourier}
\ee
can be regarded as a generalized Fourier transform (GFT) of
the function on the group $Z^D$, $f(\vec a)$, with respect
to the non-commuting variables $\vec q$. The product of two
such GFT's
\be
\tilde h(\vec q)=\tilde f(\vec q)\tilde g(\vec q)=\sum_{a\in Z^D}h(\vec a)
x_{\vec q}(\vec a)
\ee
gives rise to a deformed convolution product
\be
h(\vec a)=\sum_{\vec b\in Z^D}f(\vec b)g(\vec a-\vec b)\esp{-i\alpha(\vec
a,\vec b)}
\ee
On the contrary, eq. (\ref{NC-Fourier}) shows that the GFT
of the deformed convolution product is equal to the product
of the GFT's.   However, defining the usual Fourier
transform (FT) of the function $f(\vec a)$ in terms of the
characters of the vector representations of $Z^D$ (see eq.
(\ref{esponenziale})
\be
\tilde f_V(\vec q)= \sum_{\vec a\in Z^D}f(\vec a)\chi_{\vec q}(\vec a)
\ee
we find that the FT of $h(\vec a)$, that is of the deformed
convolution product, is  the Moyal product of the FT's of
$\tilde f_V(\vec q)$ and $\tilde g_V(\vec q)$
\be
\tilde h_V(\vec q)=\sum_{\vec a\in Z^D}h(\vec a)\chi_{\vec q}(\vec a)=
\esp{i\alpha_{\mu\nu}\partial^\mu_{\vec q_1}\partial^\nu_{\vec q_2}}
\tilde f_V(\vec q_1)\tilde g_V(\vec q_2)\Big|_{\vec q_1=\vec q_2=\vec q}
\equiv \tilde f_V(\vec q)\star\tilde g_V(\vec q)
\ee
This is a strong indication for the use of the GFT in the
harmonic analysis on the noncommuting torus. This analysis
is completed by the use of the integration theory on generic
algebras that we have introduced in \cite{casalbuoni}, and
discussed in \cite{casalbuoni2} in the case of the
projective group algebras. In fact this  theory allows us to
invert the GFT. Following ref. \cite{casalbuoni2} one
has the integration formula (depending only on the algebraic
structure and not from its representation)
\be
\int_{(x)} x(\vec a)=\delta_{\vec a,\vec 0}
\ee
which generalizes the integration over the coordinates $\vec q$ of the
commuting torus, giving
\be
\int_{(x)}\tilde f(\vec q)x(-\vec a)=f(\vec a)
\ee

In conclusion we have shown  that the  the
projective representations   of $Z^D$ are a powerful and simple tool
to find the solutions to the M
theory compactification conditions on a noncommutative
torus and to study its geometrical properties.
Also, the present methods can be easily extended to other
compactification geometries.
\vskip .3cm\noindent
{\it Note added in proof}: After  this work was completed
the paper \cite{ho} appeared in hep-th, which also
emphasizes the relevance of the projective representations
in the study of the compactification in the framework of the
M theory.


\begin{center}
{\bf Aknowledgements}
\end{center}
\medskip
The author would like to thank Joaquim Gomis and Jaume Gomis
for many enlightening discussions on the subject of
compactification. Also, he would like to thank
 Prof. J. P. Eckmann, Director of the Department of Theoretical
  Physics of the University
of Geneva, for the very kind
hospitatlity.

\end{document}